\documentclass[12pt]{iopart}
\usepackage{amsfonts,amssymb,bm,youngtab}
\usepackage{amssymb}
\usepackage{epsfig}
\usepackage{subfigure}
\usepackage{graphicx}
\usepackage{color}
\usepackage{todonotes}

\newcommand*{\defeq}{\mathrel{\vcenter{\baselineskip0.5ex \lineskiplimit0pt
      \hbox{\scriptsize.}\hbox{\scriptsize.}}}%
      =}

\usepackage{citesort}

\def\beq{\begin{equation}}
\def\eeq{\end{equation}}
\def\beqa{\begin{eqnarray}}
\def\eeqa{\end{eqnarray}}
\def\hf{\textstyle{1\over2}}
\def\3hf{\textstyle{\frac{3}{2}}}
\def\half{{1\over2}}

\newcommand{\ket}[1]{\vert #1 \rangle}

\newcommand{\bra}[1]{\langle #1 \vert}
\newcommand\unit{\mathinner{\hbox{1}\mkern-4mu\hbox{l}}}

\newcommand\ip[2]{\langle #1\,\vert\,#2\rangle}

\renewcommand{\e}{\hbox{\rm e}}
\newcommand{\myfrac}[2]{\leavevmode\kern.1em\raise.5ex\hbox{\scriptsize
$#1$}\kern-.1em {\scriptsize
/}\kern-0.10em\lower.25ex\hbox{\scriptsize $#2$}}

\def\sugroup{\hbox{SU}}

\begin{document}

\title[$D$-functions and immanants of unitary matrices and submatrices]{$D$-functions and immanants of unitary matrices and submatrices}
\author{Hubert~de~Guise$^1$, Dylan Spivak$^1$, Justin Kulp$^1$ and Ish Dhand$^2$}

\address{$^1$ Department of Physics, Lakehead University, Thunder
Bay, Ontario P7B 5E1, Canada}

\address{$\,^2$ Institute for Quantum Science and Technology, University of Calgary, Calgary, Alberta T2N 1N4 , Canada }

\begin{center}
\today
\end{center}

\begin{abstract}
Motivated by recent results
in multiphoton interferometry, 
we expand a result of Kostant on immanants of an arbitrary $m\times m$ unitary matrix $T\in \sugroup{(m)}$ to the submatrices of $T$.
Specifically, we show that immanants of principal
submatrices of a unitary matrix $T$ are a sum $\sum_{t} D^{(\lambda)}_{tt}(\Omega)$ of the diagonal $D$-functions of group element $\Omega$, 
with $t$ determined {by} the choice of submatrix, and the irrep $(\lambda)$ determined by the immanant under consideration. 
We also provide evidence that this result extends to some submatrices that are not principal diagonal, and we discuss how this result can be extended to cases where $T$ carries an $\sugroup{(m)}$ representation that is different from the defining representation.
\end{abstract}

\section{Introduction and basic result}
The immanant~$\hbox{Imm}^{\{\tau\}}(T)$ of the matrix $m\times m$ matrix 
$T$, associated with the partition $\{\tau\}$ of $S_{m}$, is defined in~\cite{Littlewood1950} by 
\beqa
\hbox{Imm}^{\{\tau\}}(T)&\defeq&\sum_{\sigma}\chi^{\{\tau\}}(\sigma)P(\sigma)\left[T_{11}T_{22}
\ldots T_{mm}\right]\, ,
\nonumber \\
&\equiv& \sum_{\sigma}\chi^{\{\tau\}}(\sigma)T_{1\sigma(1)}T_{2\sigma(2)}\ldots T_{m\sigma(m)}\, ,
\label{defineimmanant} 
\eeqa
where 
$\sigma\in S_m$ permutes $k$ to $\sigma(k)$, and $\chi^{\{\tau\}}(\sigma)$ is the character of $\sigma$ in the irrep~$\{\tau\}$ of~$S_m$.
In this paper we explore the connection between immanants and group functions (or $D$-functions) for the unitary groups, and 
extend a result of Kostant~\cite{Kostant1995} to submatrices of the fundamental representation of these groups. 

Our work is motivated in part by {intense} renewed interest in immanants of submatrices of unitary matrices and, in 
particular, of permanants~\cite{Minc1984} of submatrices and their use
in multiphoton interferometry~\cite{Tan2013,Guise2014,Tillmann2014,Shchesnovich2015,Tichy2015}.
Thus, our results are interesting within the paradigm of the BosonSampling problem~\cite{Aaronson2013,Aaronson2014} 
and its deep link to issues in computational complexity theory.

Immanants of totally non-negative and of Hermitian matrices have been studied in~\cite{Pate1992,Pate1994,Rhoades2005};
our results instead are applicable to unitary matrices and depend on the well-known
duality between representations of the unitary and of the symmetric groups~\cite{Weyl1950,Rowe2012}.

This duality identifies irreps of {SU$(m)$ with irreps of $S_N$ with $N\le m$.}
If  $\{\lambda\}=\{\lambda_1,\lambda_2,\ldots,\lambda_N\}$ is a partition of $\lambda=\sum_k\lambda_k$ labelling
an irrep of $S_N$, we {choose to} label irreps of SU$(m)$ using the round backets
$(\lambda)$ with {$m-1$} entries defined by {$(\lambda_1-\lambda_2,\ldots,\lambda_{N-1}-\lambda_N)$, and trailing zeroes
omitted.}  Thus, the irrep $\{21\}$ of {$S_3$} corresponds to the SU$(4)$ irrep $(110)\sim(11)$, the SU$(5)$ irrep $(1100)\sim(11)$ etc.

Group functions (or Wigner $D$-functions) are defined as the overlap between two basis states of the same irrep of SU$(m)$, one of which 
has been transformed by an element $\Omega\in \sugroup{(m)}$  . 
{If $\ket{\psi^{(\lambda)}_\ell}$, $\ket{\psi^{(\tau)}_t}$ are any two
basis states in irreps $(\lambda)$ and $(\tau)$ respectively and $T^{(\lambda)}(\Omega)$ is the matrix representing element
$\Omega\in\hbox{SU}(m)$ in the irrep $(\lambda)$}, then
\beq
D^{(\lambda)}_{\ell t}(\Omega)\defeq\bra{\psi^{(\lambda)}_\ell}T^{(\lambda)}(\Omega)\ket{\psi^{(\tau)}_t}\delta_{\lambda,\tau}\, .
\eeq

Kostant~\cite{Kostant1995} has shown a simple connection between immanants (defined in Eq.~(\ref{defineimmanant})) of the fundamental representation 
$T$ of  SU$(m)$ group elements and group functions $D^{(\lambda)} _{tt}$ of SU$(m)$ 
with $t$ running over each of the zero-weight states in irrep $(\lambda)$.
Specifically, let $\Omega\in\hbox{SU}(m)$ and $T(\Omega)$ {(no superscript)} be the defining $m\times m$ representation of $\Omega$.  
{Further define the matrix ${\cal D}^{(\tau)}(\Omega)$ by}
\beq
\left({\cal D}^{(\tau)}(\Omega)\right)_{rt}=D^{(\tau)}_{rt}(\Omega) \label{calD}
\eeq
with $r,t$ {restricted to} labelling zero-weight states in the irrep $(\tau)$.   Then we have~\cite{Kostant1995}
\beq
\hbox{Imm}^{\{\tau\}}(T(\Omega))
=\hbox{Tr}\left[{\cal D}^{(\tau)}(\Omega)\right].\label{kostantD}
\eeq

For SU$(2)$, this result simply states that the permanent of the matrix
\beq
T(\Omega)=\left(
\begin{array}{cc}
 \e^{-(\frac{1}{2}) i (\alpha +\gamma )} \cos \left(\frac{\beta }{2}\right) & -\e^{-(\frac{1}{2}) i (\alpha -\gamma )} \sin \left(\frac{\beta }{2}\right) \\
 \e^{(\frac{1}{2}) i (\alpha -\gamma )} \sin \left(\frac{\beta }{2}\right) & \e^{(\frac{1}{2}) i (\alpha +\gamma )} \cos \left(\frac{\beta }{2}\right) \\
\end{array}
\right)\, ,
\eeq
where $\Omega=(\alpha,\beta,\gamma)\in \hbox{SU}(2)$ is the SU$(2)$ function $\hbox{Imm}^{\{2\}}(T(\Omega))=D^1_{00}(\alpha,\beta,\gamma)=\cos\beta$ while
the determinant $\hbox{Imm}^{\{1,1\}}(T(\Omega))=D^0_{00}(\alpha,\beta,\gamma)=1$. 
The trace of Eq.~(\ref{kostantD}) contains a single term in both SU$(2)$ cases as the zero-weight subspaces in irreps $J=1$ and $J=0$ are both one-dimensional.
Here and henceforth we follow the physics notation of
labelling SU$(2)$ irreps using the angular momentum label $J=\hf\lambda$, such that $2J$ is an integer. Thus, $D^{1}_{00}(\Omega)$ is an
SU$(2)$ $D$-function in the three-dimensional irrep $J=1$.

\section{Notational details}

We first introduce a basis for ${\Bbb H}^{(1)}_p$, which is the $p$'th copy of the
carrier space for fundamental irrep $\{1\} \equiv (1)$ of SU$(m)$. We write this basis in terms of harmonic oscillator states according to  
\beq
{\Bbb H}^{(1)}_p=\hbox{Span}\{a_k^\dagger(\omega_p)\ket{0}\, , k=1,\ldots,m\}\, .
\eeq
The label $\omega_p$ can be thought of as an internal degree of freedom, say the frequency, of the $p$'th oscillator.
We introduce the (reducible) Hilbert space 
${\Bbb H}^{(N)} \defeq {\Bbb H}^{(1)}_1\otimes {\Bbb H}^{(1)}_2\ldots\otimes {\Bbb H}^{(1)}_N$, which is spanned by the set of harmonic  oscillator states of the type
\beq
a_k^\dagger(\omega_1)a_r^\dagger(\omega_2)\ldots a_s^\dagger(\omega_N)\ket{0}\, , \quad
k =1,\ldots,m\, ;\quad r=1,\ldots,m\, ,{\it etc}\label{Hpstates}
\eeq

Another ingredient we need is the action of the permutation group $S_N$ on ${\Bbb H}^{(N)}$.
The action of $P(\sigma)$ is defined as
\beqa
&&P(\sigma)a_k^\dagger(\omega_1)a_r^\dagger(\omega_2)\ldots a_s^\dagger(\omega_N)\ket{0}\nonumber \\
&&\qquad\qquad =a_k^\dagger(\omega_{\sigma^{-1}(1)})a_r^\dagger(\omega_{\sigma^{-1}(2)})\ldots a_s^\dagger(\omega_{\sigma^{-1}(N)})\ket{0}. 
\label{leftaction}
\eeqa
Alternatively, one may consider each of the sets $\{a^\dagger_k(\omega_p); k=1,\ldots,m\}$, which are labelled by $p$, as a tensor operator that carries the defining irrep $(1)$ of $\sugroup(m)$.
$\ket{0}$ is invariant under the action of $S_N$ and $\sugroup(m)$ elements.

The algebra u$(m)$ is spanned by the $S_N$-invariant operators
\beq
\hat C_{ij}=\sum_{k=1}^N a_i^\dagger(\omega_k)a_j (\omega_k)\quad i,j=1,\ldots, m.
\eeq
The su$(m)$ subalgebra is obtained from u$(m)$ by removing the diagonal operator $\sum_{i=1}^m\hat C_{ii}$,
so the Cartan subalgebra of su$(m)$ is spanned by the traceless diagonal operators 
\begin{equation}
\hat h_i \defeq \hat C_{ii}-\hat C_{i+1,i+1},\quad i=1,\ldots, m-1. 
\end{equation}

A basis for the irrep $(\lambda)$ of $\sugroup(m)$ is given in terms of the harmonic oscillator occupation number $n$ according to
\begin{equation}
\ket{\psi^{(\lambda)}n;\Lambda}=
\ket{(\lambda)n_1n_2,\ldots,n_m;(\lambda')\ldots (J)}, 
\end{equation} 
where $n\defeq(n_1,n_2,\ldots,n_m)$ and $n_k$ indicates the number of excitations in mode $k\le m$.
The weight of this state is equals the $(m-1)$-tuple $[n_1-n_2,n_2-n_3,\ldots, n_{m-1}-n_m]$. 
Finally, the multi-index $\Lambda\defeq(\lambda')\ldots (J)$ refers to a collection of indices, each of which labels irreps in 
the subalgebra chain
\beq
\begin{array}{cccccc}
\hbox{su}(m)&\supset& \hbox{su}(m-1)&\supset&\cdots &\supset \hbox{su}(2)\\
(\lambda) && (\lambda') && & (J) 
\end{array},
\label{subalgebrachain}
\eeq
and is needed to fully distinguish states having the same weight. The representation labels are all integers, except as mentioned above for 
the half-integered $\hbox{SU}(2)$ label $J$.
We take the subalgebra
$\hbox{su}(k-1)\subset \hbox{su}(k)$ to be spanned by the subset of the $k\times k$ hermitian traceless matrices of the form
\beq
\left(\begin{array}{cccc}
0&0&\ldots&0\\
0&*&*&*\\
\vdots &*&*&*\\
0&*&*&*\end{array}\right),
\eeq
where $*$ denote possible non-zero entries in $\hbox{su}(k-1)$. 

Using this notation, we can write the matrix representation of $\Omega\in\hbox{SU}(3)$ in the fundamental representation 
(which is denoted by (1) once the trailing $0$ has been eliminated)  of SU$(3)$ as
\beq
T(\Omega)=\left(\begin{array}{ccc}
D^{(1)}_{100(0);100(0)}(\Omega)&D^{(1)}_{100(0);010(\frac{1}{2})}(\Omega)&D^{(1)}_{100(0);001(\frac{1}{2})}(\Omega)\\
D^{(1)}_{010(\frac{1}{2});100(0)}(\Omega)&D^{(1)}_{010(\frac{1}{2});010(\frac{1}{2})}(\Omega)&D^{(1)}_{010(\frac{1}{2});001(\frac{1}{2})}(\Omega)\\
D^{(1)}_{001(\frac{1}{2});100(0)}(\Omega)&D^{(1)}_{001(\frac{1}{2});010(\frac{1}{2})}(\Omega)&D^{(1)}_{001(\frac{1}{2});001(\frac{1}{2})}(\Omega)
\end{array}\right)\, ,
\eeq
with $\Omega\in \hbox{SU}(3)$. The result of Kostant~\cite{Kostant1995} applied to $\sugroup{(3)}$ then states that
\beqa
\hbox{Per}(T(\Omega))&=&\hbox{Imm}^{\{3\}}(T(\Omega))=D^{(3)}_{111(1);111(1)}(\Omega)\, ,\nonumber \\
&&\hbox{Imm}^{\{21\}}(T(\Omega))=D^{(11)}_{111(1);111(1)}(\Omega)
+D^{(11)}_{111(0);111(0)}(\Omega)\, ,\\
\hbox{Det}(T(\Omega))&=&\hbox{Imm}^{\{111\}}(T(\Omega))=D^{(0)}_{000(0);000(0)}(\Omega)=1\, ,\nonumber
\eeqa
For convenience, we lighten notation and use
the symbols $T$ and $\Omega$ to respectively denote matrices and elements in different SU$(m)$ without
indicating $m$; this does not affect our conclusions as our results apply to any $m$.

The novel contribution of this paper is to extend result of~\cite{Kostant1995} encapsulated in Eq.~(\ref{kostantD}) 
to submatrices of the fundamental representations. In addition, using
outer plethysms, we can extend the theorem to immanants of matrix representations of SU$(m)$ beyond the fundamental representation and to matrix representations of subgroups
of SU$(m)$.

\section{Proving the theorem: {the case $N=m$}}

We consider the state $\ket{\Psi_{123\ldots m}}=a_1^\dagger(\omega_1)a_2^\dagger(\omega_2)\ldots a_m^\dagger(\omega_m)\ket{0}$, which lives in the (reducible) tensor product space ${\Bbb H}^{(m)}={\Bbb H}^{(1)}_1\otimes {\Bbb H}^{(1)}_2\ldots\otimes {\Bbb H}^{(1)}_m$.
The first lemma deals with the weight of this state.
\medskip

\noindent\textbf{Lemma 1:} 
The SU$(m)$ weight of $\ket{\Psi_{123\ldots m}}$ is [0]. 
\medskip

\noindent \textit{Proof}: 
 This is immediate since every mode is occupied once, so $n_i=1 \forall\, i$. Since the component $k$ of the 
 weight is $n_k-n_{k+1}$, $\hat h_i\ket{\Psi_{123\ldots m}}=0\  \forall\, i$.\hfill $\blacksquare$\medskip

From Lemma~1, we write $\ket{\Psi_{123\ldots m}}$ as an expansion over zero-weight states in all irrep occurring in ${\Bbb H}^{(m)}$ according to
\beq
\ket{\Psi_{123\ldots m}}=\sum_{\alpha\lambda\ell}\tilde c^{(\lambda)_\alpha}_{\ell} \ket{\psi^{(\lambda)_\alpha}_{\ell}}\, ,\qquad
\tilde c^{(\lambda)_\alpha}_{\ell}=\ip{\psi^{(\lambda)_\alpha}_{\ell}}{\Psi_{123\ldots m}}\, ,
\eeq
where $(\lambda)_\alpha$ is the $\alpha$'th copy of the irrep $(\lambda)_\alpha$ of SU$(m)$
and $\ell$ labels those basis states that have $0$-weight in the irrep $(\lambda)_\alpha$ of SU$(m)$.

\medskip

\noindent\textbf{Lemma 2:} With the notation above:
\beq
\sum_{\sigma}\chi^{\{\tau\}}(\sigma)P(\sigma)\ket{\Psi_{123\ldots m}}=\frac{m!}{\dim(\tau)}
\sum_{\alpha t}\tilde c^{(\tau)_\alpha}_{t}\, \ket{\psi^{(\tau)_\alpha}_{t}}.
\eeq

\medskip

\noindent \textit{Proof}: 
The proof of Lemma~2 relies on the duality between representations of the symmetric and the unitary groups.
From duality, the basis states $\{\ket{\psi^{(\tau)_\alpha}_{t}}\}$ are also basis states for the irrep $\{\tau\}$ of $S_m$. Hence, using
Eq.~(\ref{leftaction}) we obtain
\beqa
P(\sigma)\ket{\Psi_{123\ldots m}}&=&
\sum_{\alpha\lambda\ell}
\ket{\psi^{(\lambda)_\alpha}_{\ell}}\bra{\psi^{(\lambda)_\alpha}_{\ell}}P(\sigma)\ket{\Psi_{123\ldots m}}\, ,\\
&=&\sum_{\alpha\ell\lambda k} \ket{\psi^{(\lambda)_\alpha}_{\ell}}
\Gamma^{\{\lambda\}}_{\ell k}(\sigma)
\ip{\psi^{(\lambda)_\alpha}_{k}}{\Psi_{123\ldots m}}\, ,\\
&=&\sum_{\alpha\ell\lambda k} \ket{\psi^{(\lambda)_\alpha}_{\ell}}
\Gamma^{\{\lambda\}}_{k\ell}(\sigma^{-1})\,\tilde c^{(\lambda)_\alpha}_{ k},
\eeqa
where $\Gamma^{\{\lambda\}}$ is the unitary irrep $\{\lambda\}$ of $S_m$. 
Writing $\chi^{\{\tau\}}(\sigma)=\sum_{t}\Gamma^{\{\tau\}}_{tt}(\sigma)$ gives us
\beqa
\sum_{\sigma}\chi^{\{\tau\}}(\sigma)P(\sigma)\ket{\Psi_{123\ldots m}}&=&
\sum_{\alpha k\lambda\ell}\tilde c^{(\lambda)_\alpha}_{\ell} \ket{\psi_\ell^{(\lambda)_\alpha}}
\left[\sum_{\sigma t}\Gamma^{\{\tau\}}_{tt}(\sigma)\Gamma^{\{\lambda\}}_{\ell k}(\sigma^{-1})\right],\\
&=&\frac{m!}{\dim(\tau)}\sum_{\alpha t}\tilde c^{(\tau)_\alpha}_{t} \ket{\psi^{(\tau)_\alpha}_{t}}\, ,\label{Eq:mFactDimensions}
\eeqa
where we have used the orthogonality of characters to arrive at Eq.~(\ref{Eq:mFactDimensions}). \hfill$\blacksquare$\medskip

Because the action of $\Omega\in\hbox{SU}(m)$ commutes with the action of 
$\sigma\in S_m$, we have
\beqa
\hbox{Imm}^{\{\tau\}}(T(\Omega))&=&\sum_{\sigma}\,\chi^{\{\tau\}}(\sigma)P(\sigma)
\left[T_{11}(\Omega)T_{22}(\Omega)\ldots T_{mm}(\Omega)\right]\, ,\\
&=&\bra{\Psi_{123\ldots m}}{\left[T(\Omega)\otimes T(\Omega)\ldots\otimes T(\Omega)\right]} \nonumber \\
&&\qquad\qquad\left[\sum_{\sigma}\,\chi^{\{\tau\}}(\sigma)P(\sigma)\right] \ket{\Psi_{123\ldots m}}\, ,\\
&=&\sum_{\alpha rt} (\tilde c^{\tau_\alpha}_{r})^*\,\tilde c^{\tau_\alpha}_{t}\,\frac{m!}{\hbox{dim}(\tau)}D^{(\tau)}_{rt}(\Omega)\, .
\eeqa
Introducing the scaled coefficients 
$c^{(\tau)_\alpha}_{t}= \tilde c^{(\tau)_\alpha}_{t} \sqrt{\displaystyle\frac{m!}{\hbox{dim}(\tau)}}$,
we finally obtain
\beq
\hbox{Imm}^{\{\tau\}}(T(\Omega))=\sum_{rt} \left[\sum_\alpha (c^{(\tau)_\alpha }_r)^*\, c^{(\tau)_\alpha}_{t}\right]\,D^{(\tau)}_{rt}(\Omega)\, ,
\label{mixedsum}
\eeq
where the sums over $t$ and $r$ is a sum over zero-weight states in $(\tau)_\alpha$.

This result is not unexpected as the operator
\beq
\hat \Pi^{\{\tau\}}=\left[\sum_{\sigma}\chi^{\{\tau\}}(\sigma)P(\sigma)\right] \, ,\qquad \sigma\in S_m\label{immanantoperator}
\eeq
is a projector to that subspace of $S_m$ which has permutation symmetry $\{\tau\}$, and 
hence (by duality) is a projector to a subspace that carries (possibly multiple copies of) the irrep $(\tau)$ of SU($m$) 
in the $m$-fold product $(1)^{\otimes m}$.

\medskip

\noindent\textbf{Theorem 3:}~(Kostant~\cite{Kostant1995}) $\hbox{Imm}^{\{\tau\}}(T(\Omega))=\sum_{t}D^{(\tau)}_{tt}(\Omega)$.
\medskip

\noindent \textit{Proof}: We present a proof that will {eventually} allow us to dispense with the requirements that 
{$N=m$ and that} states have zero-weight. Construct the matrix 
\beq
W^{\{\tau\}}_{rt}= \sum_{\alpha} c^{(\tau)}_{\alpha t}\, (c^{(\tau)}_{\alpha r})^*\, .
\eeq
Eq.~(\ref{mixedsum}) then becomes
\beq
\hbox{Imm}^{\{\tau\}}(T(\Omega))=\sum_{rt} W^{\{\tau\}}_{rt}\,D^{(\tau)}_{rt}(\Omega)=\hbox{Tr}\left[W^{\{\tau\}}{\cal D}^{(\tau)}(\Omega)\right]\, ,
\label{tracesum}
\eeq
with ${\cal D}^{(\tau)}(\Omega)$ defined in Eq.~(\ref{calD}). Our objective is to prove that $W^{\{\tau\}}$ is the unit matrix.

Any immanant has the property of invariance
under conjugation by elements {in $S_m$} {\it i.e.}, the immanant of any matrix satisfies
\beqa
\hspace{-1cm}\hbox{Imm}^{\{\tau\}}(T(\Omega))&=\sum_\sigma \chi^{\{\tau\}}(\sigma) P(\sigma) \,\left[T_{11}(\Omega)T_{22}(\Omega)\ldots T_{mm}(\Omega)\right]
\, \nonumber \\
\qquad &=\sum_\sigma \chi^{\{\tau\}}(\sigma) P^{-1}(\bar\sigma)P(\sigma) P(\bar\sigma)\,\left[T_{11}(\Omega)T_{22}(\Omega)\ldots T_{mm}(\Omega)\right]
\, ,
\eeqa
with $\sigma,\bar\sigma \in S_m$. Under conjugation by $\bar\sigma$, Eq.~(\ref{mixedsum}) becomes
\beq
\begin{array}{rcl}
\hbox{Imm}^{\{\tau\}}(T(\Omega))
&=& {\hbox{Tr}\left[\Gamma^{\{\tau\}}(\bar \sigma) W^{\{\tau\}} \Gamma^{\{\tau\}}(\bar \sigma^{-1}){\cal D}^{(\tau)}(\Omega)\right]}\, , \\
&=&{\hbox{Tr}\left[W^{\{\tau\}} {\cal D}^{(\tau)}(\Omega)\right] }\, .
\end{array}\, .
\eeq
{Since ${\cal D}^{(\tau)}(\Omega)$ is certainly \emph{not} the unit matrix for arbitrary $\Omega$}, it follows that 
\beq
{\Gamma^{\{\tau\}}(\bar \sigma) W^{\{\tau\}} \Gamma^{\{\tau\}}(\bar \sigma^{-1})= W^{\{\tau\}},}
\eeq
{\it i.e.} the matrix $W^{\{\tau\}}$ is invariant under any permutation. By Schur's lemma 
$W^{\{\tau\}}$ must therefore be proportional to the unit matrix, {\it i.e.}~we have
$W^{\{\tau\}}_{ts}=\xi\, \delta_{ts}$ with $\xi $ the relevant constant of proportionality. 
The immanant thus takes the form
\beq
\hbox{Imm}^{\{\tau\}}(T(\Omega))=\xi\left(\displaystyle\sum_{t}D^{(\tau)}_{tt}(\Omega)\right)\, .
\eeq
To determine $\xi$, choose $\Omega=\unit$. Then $T(\unit)$ is the $m\times m$ unit matrix, and 
$T_{k,\sigma(k)}(\unit)$ is zero unless $\sigma=\unit\in S_m$. The immanant for $\Omega=\unit$ is then just the dimension
of the irrep $\{\tau\}$ and we have
\beq
\hbox{Imm}^{\{\tau\}}(T(\unit))=\chi^{\{\tau\}}(\unit)=
\hbox{dim}(\tau),\quad\xi=\left(\sum_{t} 1\right)=\xi\,\hbox{dim}(\{\tau\})
\eeq
since $D^{(\tau)}_{tt}(\unit)=1$. Hence, $\xi=1$ and the theorem is proved.\hfill $\blacksquare$\medskip

\section{Results on submatrices: {the case $N < m$.}}

We now consider the submatrices of $T$. In multiphoton interferometry, such submatrices describe the unitary scattering from an input state of the form 
\beq
\ket{\Psi_{k_1\ldots k_p}}=a^\dagger_{k_1}(\omega_1)a^\dagger_{k_2}(\omega_2)\ldots a^\dagger_{k_p}(\omega_p)\ket{0}\, , 
\qquad p< m\, , \label{substate}
\eeq
to an output state $\ket{\Psi_{\ell_1\ldots \ell_p}}$, which need not the identical to $\ket{\Psi_{k_1\ldots k_p}}$.  
{Both input and output
live in the reducible Hilbert space ${\Bbb H}^{(p)}$, and have expansions of the form
\beq
\ket{\Psi_{k_1\ldots k_p}}=\sum_{\alpha \lambda \ell} \tilde d^{(\lambda)_{\alpha}}_\ell \ket{\psi^{(\lambda)_{\alpha}}_\ell}
\label{expandgeneral}
\eeq
where $\ket{\psi^{(\lambda)_{\alpha}}_\ell}$ has weight $[k_1-k_2,k_2-k_3,\ldots,k_{p-1}-k_p]$.}

\subsection{Principal coaxial submatrices}

First we select from $T(\Omega)$ a principal submatrix $\bar T(\Omega)_k$, {\it i.e.,}~$\bar T(\Omega)_k$ is obtained by keeping rows and columns $k=(k_1,k_2,\ldots,k_p)$
with $p< m$. In such a case, the input and output states are identical.  
The permutation group $S_p$ shuffles the $p$ indices $k_1,k_2,\ldots,k_p$ amongst themselves. 
Although the submatrix 
$\bar T(\Omega)_k$ is not unitary, the proof of Theorem 3 does not depend on the unitarity of 
$T(\Omega)$ and so can be copied to show

\medskip
\noindent\textbf{Corollary 4:} The immanant 
$\hbox{Imm}^{\{\lambda\}}_k(T(\Omega))$ of a submatrix $\bar T(\Omega)_k$, which is a principal submatrix of $T$, is given by
\beq
\hbox{Imm}^{\{\lambda\}}_k(T(\Omega))=\sum_r D^{(\lambda)}_{rr}(\Omega) \label{corollary1}
\eeq
where $(\lambda)$ is the irrep of SU$(m)$ corresponding to the partition 
$\{\lambda\}$, and where the sum over $r$ is a sum over all the states in $(\lambda)$ with weight  $[k_1-k_2,k_2-k_3,\ldots,k_{p-1}-k_p]$; 
{following Eq.(\ref{expandgeneral}) this is the weight of $\ket{\Psi_{k_1\ldots k_p}}$ in Eq.~(\ref{substate})}
and need not be zero.\hfill$\blacksquare$\medskip

As an example, one can verify that, if we strip the $5\times 5$ fundamental matrix representation of SU$(5)$ from its third 
and fifth rows and columns, then
the states {entering} in the sum of Eq.~(\ref{corollary1}) are {linear combinations of terms of} the form 
\beq
P(\sigma)\left[a^\dagger_{1}(\omega_1)a^\dagger_{2}(\omega_2) a^\dagger_{4}(\omega_3)\ket{0}\right]
\eeq
with weight $[0,1,-1,1]$. 
Using the $\hbox{su}(k)\downarrow\hbox{su}(k-1)$ branching rules~\cite{Slansky1981,Whippman1965} to label basis states,  
the $\{2,1\}$ immanant of this submatrix is the sum
\beqa
\hbox{Imm}^{\{2,1\}}_{124}(T(\Omega))&&=D^{(1,1)}_{11010(2)(1)(\half);11010(2)(1)(\half)}(\Omega) \nonumber \\
 &&\quad +D^{(1,1)}_{11010(0,1)(1)(\half);11010(0,1)(1)(\half)}(\Omega)\, ,
\eeqa
where the labels $(2)(1)(\half) $ and $(0,1)(1)(\half)$ refer to the $\hbox{su}(4)\supset\hbox{su}(3)\supset\hbox{su}(2)$ chains of irreps
(recall that trailing 0s are omitted).

Littlewood~\cite{Littlewood1950} has established a number of relations between immanants of a matrix and sums of products of immanants of 
principal coaxial submatrices.  For instance, {the equality for Schur functions $\{3\}\{1\}=\{3,1\}+\{4\}$ yields the immanant relation} 
\beqa
&&\sum_{ijk\ell} \left(\hbox{Imm}^{\{3\}}_{ijk}(T(\Omega))\right)
\left(\hbox{Imm}^{\{1\}}_{\ell}(T(\Omega))\right)\nonumber \\
&&\qquad\qquad =\hbox{Imm}^{\{3,1\}}(T(\Omega))+\hbox{Imm}^{\{4\}}(T(\Omega))
\eeqa
where the sum over $ijk\ell$ is a sum over complementary coaxial submatrices, i.e.
\beq
\begin{array}{cc||cc}
ijk&\ell &ijk&\ell \\
\hline
123&4&124&3\\
134&2&234&1\\
\end{array}.
\eeq
This expands to a sum of products of immanants of submatrices given explicitly by
\beqa
&&\left(\hbox{Imm}^{\{3\}}_{123}(T(\Omega))\right)\left(\hbox{Imm}^{\{1\}}_{4}(T(\Omega))\right)+
\left(\hbox{Imm}^{\{3\}}_{124}(T(\Omega)\right) \left(\hbox{Imm}^{\{1\}}_{3}(T(\Omega))\right)\nonumber \\
&&+
\left(\hbox{Imm}^{\{3\}}_{134}(T(\Omega))\right)\left(\hbox{Imm}^{\{1\}}_{2}(T(\Omega))\right)+\left(\hbox{Imm}^{\{3\}}_{234}(T(\Omega))\right)\left(\hbox{Imm}^{\{1\}}_{1}(T(\Omega))\right)\nonumber \\
&&=\hbox{Imm}^{\{3,1\}}(T(\Omega))+\hbox{Imm}^{\{4\}}(T(\Omega)),
\eeqa
which becomes an equality on the corresponding products of sum of SU$(4)$ $D$-functions:
\beqa
&&D^{(3)}_{1110(2)(\frac{1}{2});1110(2)(\frac{1}{2})}(\Omega)D^{(1)}_{0001(1)(\frac{1}{2});0001(1)(\frac{1}{2})}(\Omega)
\nonumber \\
&&\quad +D^{(3)}_{1101(2)(\frac{1}{2});1101(2)(\frac{1}{2})}(\Omega)D^{(1)}_{0010(1)(\frac{1}{2});0010(1)(\frac{1}{2})}(\Omega)\nonumber \\
&&\quad +D^{(3)}_{1011(2)(1);1011(2)(1)}(\Omega)D^{(1)}_{0100(1)(0);0100(1)(0)}(\Omega)\nonumber \\
&&\quad +D^{(3)}_{0111(3)(1);0111(3)(1)}(\Omega)D^{(1)}_{1000(0)(0);1000(0)(0)}(\Omega)\nonumber \\
&&=
D^{(21)}_{1111(3)(1);1111(3)(1)}(\Omega)+D^{(21)}_{1111(11)(1);1111(11)(1)}(\Omega)
\nonumber \\
&&\quad+D^{(21)}_{1111(11)(0);1111(11)(0)}(\Omega) +D^{(4)}_{1111(3)(1);1111(3)(1)}(\Omega).
\eeqa
The subgroup labels are obtained by systematically using the su$(k)\downarrow\hbox{su}(k-1)$ branching rules~\cite{Dhand2015}.

\subsection{Generic submatrices}

To fix ideas, we start with the $4\times 4$ matrix $T$ and remove row $1$ and column $2$ to obtain the submatrix $\bar T$:
\beq
T(\Omega)\to \bar T(\Omega)=
\left(
\begin{array}{cccc}
T_{21}(\Omega)&T_{23}(\Omega)&T_{24}(\Omega)\\
T_{31}(\Omega)&T_{33}(\Omega)&T_{34}(\Omega)\\
T_{41}(\Omega)&T_{43}(\Omega)&T_{44}(\Omega)
\end{array}
\right)
\label{su4submatrix}
\eeq
The immanants of $3\times 3$ submatrix $\bar T(\Omega)$ are in the form
\beqa
\hbox{Imm}^{\{\lambda\}}(\bar T(\Omega))&=&\sum_{\sigma}\chi^{\{\lambda\}}(\sigma)P(\sigma)\left[T_{11}(\Omega)T_{22}(\Omega)T_{34}(\Omega)\right]\, \\
&=&\hat \Pi^{\{\lambda\}} \left[T_{11}(\Omega)T_{22}(\Omega)T_{34}(\Omega)\right]\, ,
\eeqa
where $\hat \Pi^{\{\lambda\}}$ is the immanant projector of Eq.~(\ref{immanantoperator}) and 
$\sigma$ permutes the triple $(124)$. 

Let $\{a_k^\dagger(\omega_k)\ket{0},k=1,\ldots,4\}$ be a basis for the fundamental irrep of SU($4$), and define
\beqa 
\ket{\Psi_{134}}&\defeq&a_1^\dagger(\omega_1)a_3^\dagger(\omega_2)a_4^\dagger(\omega_3)\ket{0}\, ,\\
\ket{\Phi_{234}}&\defeq&a_2^\dagger(\omega_1)a_3^\dagger(\omega_2)a_4^\dagger(\omega_3)\ket{0}\,
\eeqa
as three-particle states elements of ${\Bbb H}^{\{1\}\otimes\{1\}\otimes\{1\}}$.
Clearly there is $\sigma'\in S_4$ such that
\beq
\ket{\Psi_{234}}=P(\sigma')\ket{\Phi_{134}}\, .
\eeq
Indeed by inspection this element is given by $P(\sigma')=P_{12}$. More generally, if 
\beqa
\ket{\Phi_k}&=&a_{k_1}^\dagger(\omega)a_{k_2}^\dagger(\omega_2)a_{k_3}^\dagger(\omega_3)\ket{0}\, ,\qquad
k=(k_1,k_2,k_3)\, ,\\
\ket{\Psi_q}&=&a_{q_1}^\dagger(\omega)a_{q_2}^\dagger(\omega_2)a_{q_3}^\dagger(\omega_3)\ket{0}\, ,\qquad
q=(q_1,q_2,q_3)\, ,
\eeqa
then there is $\sigma_{qk}$ exists such that $\ket{\Psi_{q}}=P(\sigma_{qk})\ket{\Phi_{k}}$.  
As the action of the permutation group commutes with the action of the unitary group:
\beqa
\hbox{Imm}^{\{\lambda\}}(\bar T(\Omega))_{kq}&=&\bra{\Phi_{k}}\,\left[T(\Omega)\otimes T(\Omega)\ldots\otimes T(\Omega)\right]
\,\hat \Pi^{\{\lambda\}}\,P(\sigma_{qk})\ket{\Phi_{k}}\, ,\\
&=&\bra{\Phi_{k}}\,\hat \Pi^{\{\lambda\}}\,\left[T(\Omega)\otimes T(\Omega)\ldots\otimes T(\Omega)\right]P(\sigma_{qk})\ket{\Phi_{k}}\, , \nonumber \\
&=&\sum_{rs\alpha}\bra{\Phi_{k}}\hat \Pi^{\{\lambda\}}\ket{\psi^{(\lambda)_{\alpha}}_r} \nonumber \\
&&\qquad\times \bra{\psi^{(\lambda)_{\alpha}}_r}T^{(\lambda)}(\Omega)P(\sigma_{qk})
\ket{\psi^{(\lambda)_\alpha}_s}\ip{\psi^{(\lambda)_\alpha}_s}{\Phi_{k}} \label{immanantnondiagonal}
\eeqa
Now, the permutation $P(\sigma_{qk})$ is represented by a unitary matrix in the carrier space $(\lambda)_\alpha$.
Thus, there exist $\Omega_{qk}\in \sugroup{(4)}$ and a phase $\zeta$ such that 
$P(\sigma_{qk})\ket{\psi^{(\lambda)_\alpha}_s}=e^{i\zeta}\,T(\Omega_{qk})\ket{\psi^{(\lambda)_\alpha}_s}$.
This transforms our original problem back to the case of principal submatrices, but with now an element
$\Omega\cdot \Omega_{qk}$ {\it i.e.},
\beq
\hbox{Imm}^{\{\lambda\}}(\bar T(\Omega))=\sum_{t}D^{(\lambda)}_{tt}(\Omega\cdot \Omega_{qk})\, .
\eeq
Unfortunately, the action $P(\sigma_{qk})\ket{\psi^{(\lambda)_\alpha}_s}$ is in general highly non-trivial~\cite{Kramer1966,Moshinsky1968,Rowe1999}~and it is not obvious
how to find $\Omega_{qk}$, much less $(\Omega\cdot \Omega_{qk})$. Nevertheless, we found that the sum of $D$-functions that occur on the right hand side of 
Eq.~(\ref{immanantnondiagonal}) always contains the same number of $D$ as the dimension of the dual irrep $\{\tau\}$, and that the coefficients of these 
$D$'s is always one. This result relied on (i) evaluating the appropriate group functions using the algorithm~\cite{Dhand2015}, (ii) explicitly constructing each of 
the immanants of all possible $4\times 4$ submatrices and of all possible $3\times 3$ submatrices of the fundamental irrep of $\sugroup{(5)}$ and (iii) explicitly 
constructing the immanants of $3\times 3$ submatrices of the fundamental irrep of $\sugroup{(4)}$ or $\sugroup{(5)}$.

Thus, in the specific case of the submatrix given in Eq.~(\ref{su4submatrix}), we have 
\beq
\hspace{-1.2cm}\mathrm{Imm}^{\{21\}}(\bar T(\Omega))_{(234)(134)}=D^{(1,1)}_{0111(2)(\frac{1}{2});1011(2)(\frac{1}{2})}(\Omega) 
+ D^{(1,1)}_{0111(11)(\frac{1}{2});1011(11)(\frac{1}{2})}(\Omega)\, .
\eeq
We also verified that a similar identity holds for all $3\times 3$ submatrices of $T(\Omega)\in\sugroup{(4)}$. For instance, 
\begin{eqnarray}
\hspace{-1.2cm}\mathrm{Imm}^{\{21\}}(\bar T(\Omega))_{(234)(124)}&=&D^{(1,1)}_{0111,(2)(\frac{1}{2});1101(2)(\frac{1}{2})}(\Omega) + D^{(1,1)}_{0111(11)(\frac{1}{2});1101(01)(\frac{1}{2})}(\Omega)\, ,\\
\hspace{-1.2cm}\mathrm{Imm}^{\{21\}}(\bar T(\Omega))_{(134)(124)}&=& D^{(1,1)}_{1011(2)(\frac{1}{2});1110(2)(\frac{1}{2})}(\Omega) + D^{(1,1)}_{1011(11)(\frac{1}{2});1110(01)(\frac{1}{2})}(\Omega)\, .
\end{eqnarray}
Likewise, we have, for $T(\Omega)\in \sugroup(5)$, 
\beqa
&&\mathrm{Imm}^{\{21\}}(\bar T(\Omega))_{(1345)(1235)}\nonumber \\
&&\quad=D^{(1,1)}_{01101(11)(2)(\frac{1}{2}),10110(2)(2)(\frac{1}{2})}(\Omega)
+D^{(1,1)}_{01101(11)(01)(\frac{1}{2}),10110(01)(01)(\frac{1}{2})}(\Omega)\, ,\\
&&\mathrm{Imm}^{\{31\}}(\bar T(\Omega))_{(235)(134)}=D^{(2,1)}_{10111(3)(3)(1),11101(3)(2)\frac{1}{2}}(\Omega)\nonumber \\
&&\quad+D^{2,1}_{10111(11)(11)(1),11101(11)(2)\frac{1}{2}}(\Omega)
+D^{(2,1)}_{10111(11)(11)(0),11101(11)(01)\frac{1}{2}}(\Omega)\, ,
\eeqa
this last being an example of a $4\times 4$ submatrix not principal coaxial.
We thus conjecture that, even for generic submatrices, $\mathrm{Imm}^{\{\lambda\}}(\bar T(\Omega))_{kq}$ 
is a sum of $\hbox{dim}{\{\lambda\}}$ distinct $D$'s with coefficients equal to $+1$,
although we cannot yet formulate a solid proof.

\section{Outer plethysms and subgroups}

We now consider an application of our result to immanants of unitary matrices that are not in the fundamental representation. 
The difficulty in this case is that the various $S_N$-invariant subspaces in the $N$-fold tensor product no longer contain a single SU($m$) irrep.
Consider for instance $T(\Omega)$, the $4\times 4$ $(J=3/2)$ matrix representation of $\hbox{SU}(2)$. Using the standard
$D^J_{M'M}(\Omega)$ notation for the group functions and $\ket{JM}$ as the notation for basis states, we have
\beq
\hbox{Imm}^{\{2,2\}}(T(\Omega))=\,
\textstyle\frac{26}{35}D^4_{00}(\Omega)+\,\textstyle\frac{6}{7}D^{2}_{00}(\Omega)+\,\textstyle\frac{2}{5}D^0_{00}(\Omega)\, , 
\eeq
which is still a sum of diagonal group functions $\sum_{J}c_JD^{J}_{00}(\Omega)$. The dimension of the $S_4$ irrep 
$\{2,2\}$ is 2 and, while the sum contains more than two terms, we still have
$\sum_{J}c_J=\hbox{dim}(\{2,2\})$. The possible values of $J$ entering in the sum
are those that occur in the (outer) plethysm $(3/2)\otimes_\wp\{2,2\}$, and can be found using Schur function techniques.
(Here, $\otimes_\wp$ denotes the plethysm operation.)

A more sophisticated application is to the evaluation of immanants of the $6\times 6$ matrix representation $(2,0)$ of
SU$(3)$. 
$(2)^{\otimes 6}$ contains more than one SU$(3)$ irrep. For instance,
one might consider various immanants of the $6\times 6$ matrix $T(\Omega)$ of the $(2,0)$ of SU($3$). 
We may consider this matrix as an element
of the SU$(3)$ subgroup of SU($6)$.

As an example, the permanant  
$\hbox{Imm}^{\{6\}}(T(\Omega))=\bra{\Psi_{444}}R(\Omega)\ket{\Psi_{444}}$ 
but the state $\ket{\Psi_{444}}$ is now a linear combination of SU($3)$ states in various irreps that occur in 
$(2)\otimes_\wp \{6\}$. Explicitly, we have
\beqa
\ket{\Psi_{444}}&=&\textstyle\sqrt{\frac{64}{385}}\ket{(12)444(4)}
+\textstyle\sqrt{\frac{18}{385}}\ket{(8,2)444(\alpha)}+\textstyle\sqrt{\frac{4}{21}}\ket{(4,4)444(\beta)} \nonumber \\
&+&
\textstyle\frac{1}{3}\ket{(6)222(2)}
+\textstyle\sqrt{\frac{16}{63}}\ket{(0,6)444(2)}
+\textstyle\sqrt{\frac{2}{45}}\ket{(0)000(0)},
\label{bigpermanant}
\eeqa
where
\beqa
\ket{(8,2)444(\alpha)}&=&\textstyle\sqrt{\frac{10}{21}}\,\ket{(8,2)444(4)}+\textstyle\sqrt{\frac{11}{21}}\,\ket{(8,2)444(2)}\, 
\label{state82} \\
\ket{(4,4)444(\beta)}&=&\textstyle\sqrt{\frac{12}{35}}\,\ket{(4,4)444(4)}+\textstyle\sqrt{\frac{4}{21}}\,\ket{(4,4)444(2)}\,\nonumber \\
&&\qquad +\textstyle\sqrt{\frac{7}{15}}\,\ket{(4,4)444(0)}\, .
\label{state44}
\eeqa
Note that both $\ket{(8,2)444(\alpha)}$ and $\ket{(4,4)444(\beta)}$ carry the fully symmetric $\{3\}$ irrep of
$S_3$, as can be verified using the matrix elements of the permutation operators given in~\cite{Rowe1999}. 
The permanent $\hbox{Imm}^{\{6\}}(T(\Omega))$ can be written as a sum of diagonal $D$-functions: 
\beqa
&\hbox{Imm}&^{\{6\}}(T(\Omega))\nonumber \\
&=& c_1 D^{(12,0)}_{(444;4)(444;4)}\nonumber \\
&&+c_2D^{(8,2)}_{(444;4)(444;4)}+c_3D^{(8,2)}_{(444;4)(444;2)}+c_4D^{(8,2)}_{(444;2)(444;4)}+c_5D^{(8,2)}_{(444;2)(444;2)}\nonumber \\
&&+c_6D^{(4,4)}_{(444;4)(444;4)}+c_7D^{(4,4)}_{(444;4)(444;2)}+c_8D^{(4,4)}_{(444;4)(444;0)}+c_9D^{(4,4)}_{(444;2)(444;4)}\nonumber \\
&&+c_{10}D^{(4,4)}_{(444;2)(444;2)}+c_{11}D^{(4,4)}_{(444;2)(444;0)}+c_{12}D^{(4,4)}_{(444;0)(444;4)}+c_{13}D^{(4,4)}_{(444;0)(444;2)}\nonumber \\
&&\qquad +c_{14}D^{(4,4)}_{(444;0)(444;0)}\nonumber \\
&&+c_{15}D^{(6,0)}_{(222;2)(222;2)}+c_{16}D^{(0,6)}_{(444;2)(444;2)}+c_{17}D^{(0,0)}_{(000;0)(000;0)}
\eeqa
with solution
\beq
{\renewcommand{\arraystretch}{1.75}
\renewcommand{\arraycolsep}{3.75pt}
\begin{array}{| rl | rl | rl | rl | rl |}
\hline
c_1 =\ &\textstyle\frac{64}{385} &&&&&&&&\\
\hline
c_2=\ &\textstyle\frac{60}{539}&\ c_3=\ &\textstyle\frac{6}{49}\sqrt{\frac{10}{11}}&\ c_4=\ &\textstyle\frac{6}{49}\sqrt{\frac{10}{11}}&
c_5 =&\textstyle\frac{6}{49} &&\\
\hline
c_6=&\textstyle\frac{16}{245}&c_7=&\textstyle\frac{16\sqrt{5}}{147}&c_8=&\textstyle\frac{8}{105}&
c_9=&\textstyle\frac{16\sqrt{5}}{147}&&\\
\hline
c_{10}=&\textstyle\frac{16}{441}&c_{11}=&\textstyle\frac{8}{63\sqrt{5}}&c_{12}=&\textstyle\frac{8}{105}&
c_{13}=&\textstyle\frac{8}{63\sqrt{5}}&c_{14}=&\textstyle\frac{4}{45}\\ 
\hline
c_{15}=&\textstyle\frac{1}{9}&c_{16}=&\textstyle\frac{16}{63}&c_{17}=&\textstyle\frac{2}{45}&&&&\\
\hline
\end{array}
}
\eeq
The coefficients $c_k$ are immediately seen to be related to coefficents in Eqs.(\ref{bigpermanant})--(\ref{state44}).  
As anticipated from the theorem, the sum of coefficients
of diagonal $D$s $c_1+c_2+c_5+c_6+c_{10}+c_{14}+c_{15}+c_{16}+c_{17}=1.$

\section{Discussion and conclusion}

The relation between characters of $S_N$ and U$(n)$ is well known and a rich source of results in mathematical physics. 
Our work expands these beyond characters to novel connections between immanants and the group functions proper.
Results on group functions (see~\cite{Rowe1999,Louck1970} as well as~\cite{Dhand2015} and references therein) 
are comparatively less common than those available for, say, the calculation of generator matrix elements or Clebsch-Gordan coefficients. 
We hope some of the results given here might be useful in providing impetus or remedy to this relative paucity of results on group functions. 

Immanants are connected to the interferometry of partially distinguishable pulses~\cite{Guise2014,Tan2013,Tillmann2014}; 
the associated permutation symmetries lead to novel interpretations of 
immanants as a type of normal coordinates describing lossless passive interferometers \cite{Tillmann2014}. 
This connection immediately provides a physical interpretation to the 
appropriate combinations of group functions corresponding to these immanants, and should stimulate further development of toolkits to compute group 
functions. 
 
Conjectures in complexity theory regarding the behaviour of permanents of large unitary matrices may also provide an entry point towards understanding the behaviour of 
$D$-functions in similar asymptotic regimes. 
It remains to see if this line of thought can also be turned around: it might be possible to use results on the asymptotic 
behaviour of $D$-functions to establish some conjectures on the behaviour of immanants of large matrices.

Finally, although the Schur-Weyl duality is not directly applicable to subgroups of the unitary groups, the permutation group retains its deep connection with representations of the classical groups, which are considered as subgroups of the unitary groups~\cite{Butler1969,Wybourne1974}.
Hence, it might be possible to extend the results of this paper to 
functions of the orthogonal or symplectic groups, thus generalizing the result of Sec.~5 on immanants associated with plethysms of representations.

\section*{Acknowledgements}
This work was supported by NSERC and Lakehead University. 
ID acknowledges AITF, NSERC and USARO for financial support.
\bigskip

\section*{References}
\bibliography{ImmanantAndDs}

\end{document}